\documentclass[manuscript]{aastex}

\usepackage{graphics,color}
\usepackage{epstopdf}
\definecolor{red}{rgb}{1,0,0}

\shorttitle{PS1 Asteroid Rotation Period Survey}
\shortauthors{Chang et al.}
\begin{document}
\title{Searching for Super-Fast Rotators Using the Pan-STARRS 1}

\author{Chan-Kao Chang\altaffilmark{1}, Hsing-Wen Lin\altaffilmark{2,1}, Wing-Huen Ip\altaffilmark{1,3}, Wen-Ping Chen\altaffilmark{1}, Ting-Shuo Yeh\altaffilmark{1}, K. C. Chambers\altaffilmark{4}, E. A. Magnier\altaffilmark{4}, M. E. Huber\altaffilmark{4}, H. A. Flewelling\altaffilmark{4}, C. Z. Waters\altaffilmark{4}, R. J. Wainscoat\altaffilmark{4}, and A.S.B. Schultz\altaffilmark{4}}
\altaffiltext{1}{Institute of Astronomy, National Central University, Jhongli, Taiwan}
\altaffiltext{2}{Department of Physics, University of Michigan, Ann Arbor, Michigan 48109, USA}
\altaffiltext{3}{Space Science Institute, Macau University of Science and Technology, Macau}
\altaffiltext{4}{Institute for Astronomy, University of Hawaii, Honolulu, Hawaii 96822, USA}

\email{rex@astro.ncu.edu.tw}


\begin{abstract}
A class of asteroids, called large super-fast rotators (large SFRs), have rotation periods shorter than 2 hours and diameters larger than $\sim 0.3$ km. They pose challenges to the usual interior rubble-pile structure unless a relatively high bulk density is assumed. So far, only six large SFRs have been found. Therefore, we present a survey of asteroid rotation period using Pan-STARRS 1 telescopes during 2016 October 26 to 31 to search more large SFRs and study their properties. A total of 876 reliable rotation periods are measured, among which seven are large SFRs, thereby increasing the inventory of known large SFRs. These seven newly discovered large SFRs have diverse colors and locations in the main asteroid belt, suggesting that the taxonomic tendency and the location preference in the inner main belt of the six perviously known large SFRs could be a bias due to various observational limits. Interestingly, five out of the seven newly discovered large SFRs are mid main-belt asteroids. Considering the rare discovery rates of large SFR in the previously similar surveys \citep{Chang2015, Chang2016} and the survey condition in this work, the chance of detecting a large SFR in the inner main belt seems to be relatively low. This probably suggests that the inner main belt harbors less large SFRs than the mid main belt. From our survey, we also found the drop in number appearing at $f > 5$ rev/day on the spin-rate distribution for the outer main-belt asteroids of $D < 3$ km, which was reported for the inner and mid main belt by \citet{Chang2015, Chang2016}.
\end{abstract}


\section{Introduction}
Asteroid time-series study was a relatively unexplored field in planetary science because it was a challenge to collect a large number of asteroid light curves within a short period of time. Thanks to the significant advance in observational technology (i.e., robotic telescope and wide-field camera) and information science (i.e., high computing power and massive storage), such challenge becomes accessible and asteroid time-series study can therefore be conducted in a more comprehensive way through wide-field surveys in the last decade \citep{Masiero2009, Polishook2009, Dermawan2011, Polishook2012, Chang2014a, Chang2015, Waszczak2015, Chang2016}.

The 2-hour spin-barrier \citep{Harris1996, Pravec2002} has continuously been found for asteroids mostly with size of few hundreds meter or larger collected from these wide-field surveys. Moreover, the relation between the spin-rate limit and the bulk density of asteroids in this size range \citep[i.e., $P \sim 3.3 \sqrt{(1 + \Delta m)/\rho}$;][]{Harris1996} was first time to be seen in these data sets that the S-type asteroids has a higher spin-rate limit than the C-type asteroids \citep{Chang2015, Waszczak2015}. This suggests that the rubble-pile structure (i.e., gravitationally bounded aggregation) is generally applicable to these asteroids. However, six large super-fast rotators (large SFRs; i.e., D $> 300$ m) have been found to break the 2-hour spin-barrier and challenged the rubble-pile structure \citep[SFRs; see table 2 in ][and references therein]{Chang2017}. Although internal cohesion \citep{Holsapple2007, Sanchez2012} is a possible solution to keep these large SFRs intact under their super-fast rotations, the rarity of large SFRs, comparing to the average asteroids, somehow suggest that cohesion might be only available to certain asteroids. Moreover, a taxonomic tendency seems to present in the six known large SFRs \citep{Chang2017}. If the aforementioned rarity of large SFR and the taxonomic tendency are true, large SFRs could just possibly be a special group distinguishing from the average asteroids. Therefore, any preference shared by large SFRs, such as composition, size, and location in the main asteroid belt, is important to understand their natures.

The asteroid spin-rate distribution reflects the overall evolution of the spin state for a group of asteroids. Two dominant mechanisms, mutual collisions and the Yarkovsky-O'Keefe-Radzievskii-Paddack effect \citep[YORP;][]{Rubincam2000}, are believed to effectively alter the spin states of main-belt asteroids (MBAs). While the former (i.e., collision equilibrium) would lead to a Maxwellian spin-rate distribution \citep{Salo1987}, the latter tends to deviate the distribution from a Maxwellian form \citep{Pravec2008}. Indeed, asteroids with diameters larger than 40 km were shown to have a Maxwellian spin-rate distribution \citep{Pravec2000}, and contrarily, smaller asteroids display a distribution different from a Maxwellian form. Interestingly, some difference has been seen between the spin-rate distributions of smaller asteroids obtained from the target observations \citep[i.e., a flat distribution;][]{Pravec2008} and the wide-field surveys \citep[i.e., a deviated Maxwellian form;][]{Masiero2009, Chang2015, Waszczak2015, Chang2016} have some difference, and however, how this discrepancy was caused still needs more study. Because the timescales of both aforementioned mechanisms depend on the size and location of asteroid \citep[][and the references therein]{McNeill2016}, some footprints are therefore expected to be left in the spin-rate distributions. Fortunately, the recent wide-field surveys provide a good chance to study the spin-rate distributions of asteroids in different sizes and locations for a further insight of asteroid spin-state altering mechanisms. \citet{Chang2015, Chang2016} found that the spin-rate distributions are similar for asteroids in a fixed diameter range at different locations. Besides, a drop in number at $> 5$ rev/day was found in the spin-rate distributions for asteroids of $D < 3$ km in the inner and mid main belt, which is not seen for asteroids of $3 < D < 15$ km. The reason for this number drop is still unknown, and it is also interesting to know whether this number drop would also exist in the outer main belt.

To understand the aforementioned questions, a rotation period survey aiming at the kilometer-sized asteroids in the outer main belt is needed, and therefore, we used the Pan-STARRS1 (PS1) telescope to conduct a survey for asteroid rotation period in October 2016. From the survey, 876 reliable rotation periods were obtained and seven of them are large SFRs. The observation information and light-curve extraction are given in the Section 2. The rotation-period analysis is described in Section 3. The results and discussion can be found in Section 4, and the summary and conclusion is presented in Section 5.

\section{Observations and Data Reduction}
The Panoramic Survey Telescope And Rapid Response System-1 (Pan-STARRS1, PS1) was designed to explore the visible $3\pi$ sky and mainly dedicated to find small solar system bodies, especially those potentially hazardous objects. The telescope is a 1.8 m Ritchey–Chretien reflector located on Haleakala, Maui, which is equipped with the Gigapixel Camera \#1 to create a field of view of 7 deg$^2$. The available filters include $g_{P1}$ ($\sim 400-550$ nm), $r_{P1}$ ($\sim550-700$ nm), $i_{P1}$ ($\sim 690–820$ nm), $z_{P1}$ ($\sim820-920$ nm), and $y_{P1}$ ($>920$ nm), and a special filter, $w_{P1}$ (i.e., combination of $g_{P1}$, $r_{P1}$, and $i_{P1}$), was designed for the discovery of moving object \citep{Kaiser2010, Tonry2012, Chambers2016}.

In order to discover large SFRs and carry out the spin-rate distribution of outer MBAs down to the kilometer size, we used the PS1 to conduct a special campaign to collect asteroid light curves in $w_{P1}$ band during October 26-31, 2016, in which eight consecutive PS1 fields (i.e., $\sim56$~deg$^2$ in total) over the ecliptic plane around the opposition were continuously scanned using a cadence of $\sim10$ minutes. In the first night of the campaign, we used an observation sequence of $w_{P1}$, $g_{P1}$, $w_{P1}$, $r_{P1}$, $w_{P1}$, $i_{P1}$, $w_{P1}$, $z_{P1}$ bands to obtain asteroid colors, and the other nights were only observed in $w_{P1}$ band. The exposure times for $g_{P1}$, $r_{P1}$, $i_{P1}$, $z_{P1}$, and $w_{P1}$ bands were 120, 120, 120, 180, and 60 seconds, respectively, and this would give us a similar limiting magnitude of 22.5 mag at $5 \sigma$ level for each band. However, only few exposures were obtained for the last two nights of the campaign due to bad weather. The details of the observation can be found in Table~\ref{obs_log} and \ref{obs_log_1}.

All the images obtained in the campaign were processed by the Image Processing Pipeline (IPP), which includes image de-trending, instrumental signature removal, object detecting, image warping, and photometric and astrometric calibration \citep[the detailed description can be found in][]{Chambers2016, Magnier2016a, Magnier2016b, Magnier2016c, Waters2016}. The IPP also performs image subtraction to find transient detections and then passes them to the Pan-STARRS Moving Object Processing System to discover new moving objects \citep{Denneau2013}. From this campaign, more than 1500 asteroids were discovered and reported to the Minor Planet Center.

The light curves of asteroids, including known and newly discovered, were extracted by matching the detections against the ephemerides obtained from the {\it JPL/HORIZONS} system with a search radius of 2\arcsec ~after removing the detections of the stationary sources.

\section{Rotation-Period Analysis, Color Calculation, and Diameter Estimation}\label{period_analysis}
After correcting light-travel time and reducing both heliocentric, $r$, and geocentric, $\triangle$, distances to 1~AU for all light-curve measurements, we fitted a 2nd-order Fourier series to each light curve to find the rotation period \citep{Harris1989}:
\begin{equation}\label{FTeq}
  M_{i,j} = \sum_{k=1}^{2} B_k\sin\left[\frac{2\pi k}{P} (t_j-t_0)\right] + C_k\cos\left[\frac{2\pi k}{P} (t_j-t_0)\right] + Z_i,
\end{equation}
where $M_{i,j}$ are the reduced magnitudes in $w_{P1}$ band measured at the epoch, $t_j$; $B_k$ and $C_k$ are the coefficients in the Fourier series; $P$ is the rotation period; and $t_0$ is an arbitrary epoch. We also introduced a constant value, $Z_i$, to correct the possible offsets in magnitude between the measurements obtained from different nights. The least-squares minimization was applied to Eq.~(\ref{FTeq}) to obtain the other free parameters for each given $P$, and the explored spin-rate, $f = 1/P$, was from 0.25 to 50~rev/day with a step size of 0.01~rev/day. However, we excluded the upper and lower 5\% of the detections in a light curve in the aforementioned fitting to avoid outliers, which might be contaminated by nearby bright stars or unknown sources.

A code ($U$), describing the reliability of the derived rotation periods, was then assigned after manual review for each light curves, where `3', `2', `1', and `0' mean highly reliable, some ambiguity, possibly correct, and no detection, respectively \citep{Warner2009}. We estimated the uncertainty of rotation period using the frequency range that has $\chi^2$ smaller than $\chi_{best}^2+\triangle\chi^2$, where $\chi_{best}^2$ is the $\chi^2$ of the derived rotation period and $\triangle\chi^2$ is the 68\% (i.e., $1\sigma$) of the inverse $\chi^2$ distribution, assuming $1 + 2N_k + N_i$~degrees of freedom in which $N_k$ is the order of Fourier series and $N_i$ is the number of observation nights. The amplitude of a light curve was calculated after rejecting the upper and lower 5\% of data points.

Using the detections of different bands obtained from the first night, the colors can be calculated for the observed asteroids. To remove rotational effect in the color calculation, an offset for each band was simply fitted using Eq.~\ref{FTeq} with the solution obtained from the rotation period fitting. Therefore, only asteroids with a rotation period of $U >= 2$ have color calculation. However, we rejected a case if its detections in $g_{P1}$, $r_{P1}$, $i_{P1}$, and $z_{P1}$ bands do not well follow its folded light curve in $w_{P1}$ band. Moreover, we adopted the first order translation from \citet{Tonry2012} to covert the PS1 color into SDSS color, and then, determined the spectral type using the SDSS color, $a^*$\footnote{$a^* = 0.89*(g-r) + 0.45*(r-i)-0.57$, which was first used to distinguish blue ($a^* < 0$) and red ($a^* > 0$) asteroids in the SDSS $r-i$ vs $g-r$ diagram \citep{Ivezic2001}.} vs $i-z$ \citep{Ivezic2002}, and the boundary defined by \citet{Parker2008}\footnote{The SDSS colors of C- and X-type (i.e., including the E-, M-, and P-type) are overlapped in the region of $a^* < 0$ \citep[i.e., the neutral color objects;][]{Demeo2013}. To distinguish the C- and X-type asteroids relies on albedo or spectrum. In this work, we follow the definition of \citet{Parker2008} to show the diverse colors of our samples.}.

Since the phase angles only had a small variation during our relatively short observation time-span, a fixed $G_{w}$ slope of 0.15 in the $H$--$G$ system was simply applied to estimate the absolute magnitudes of asteroids \citep{Bowell1989}. We then estimated the diameter using
\begin{equation}\label{dia_eq}
  D = {1329 \over \sqrt{p_{v}}} 10^{-H_{w}/5},
\end{equation}
where $H_v$ is the absolute magnitude in $V$ band converted from the $H_{w}$ from our observation, $D$ is the diameter in~km, $p_v$ is the $V$ band geometric albedo, and 1329 is the conversion constant. We adopted the albedo value for S-, V-, and C-type to be $p_v$ = 0.23, 0.35, and 0.06 from \citet{Demeo2013} if the asteroid has its spectral type determination from our observation. Otherwise, three empirical albedo values, $p_w = 0.20$, 0.08 and 0.04, were assumed for asteroids in the inner ($2.1 < a < 2.5 AU$), mid ($2.5 < a < 2.8$ AU) and outer ($a > 2.8$ AU) main belts, respectively \citep{Tedesco2005}. However, if the $WISE$/$NEOWISE$ diameter estimation of an asteroid is available, we then adopted that value \citep{Grav2011, Mainzer2011, Masiero2011}.

\section{Results and Discussion}
\subsection{The Derived Rotation Periods and Colors}\label{discuss_p}
\begin{figure}
\plotone{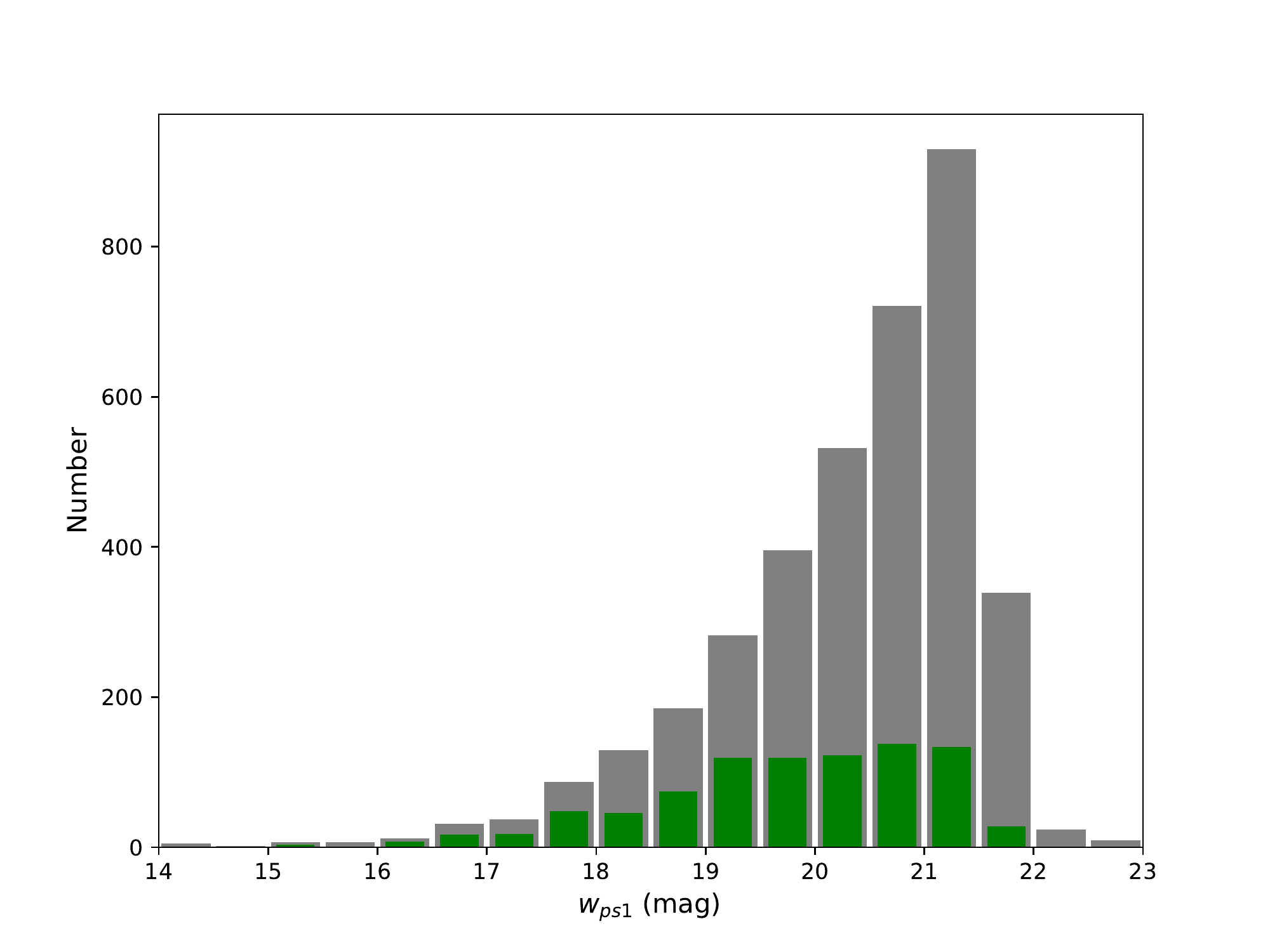}
\caption{Magnitude distribution of asteroids observed in the survey. Green: asteroids with reliable rotation periods;  gray: asteroids with 10 or more detections. }
\label{mag_hist}
\end{figure}

\begin{figure}
\plotone{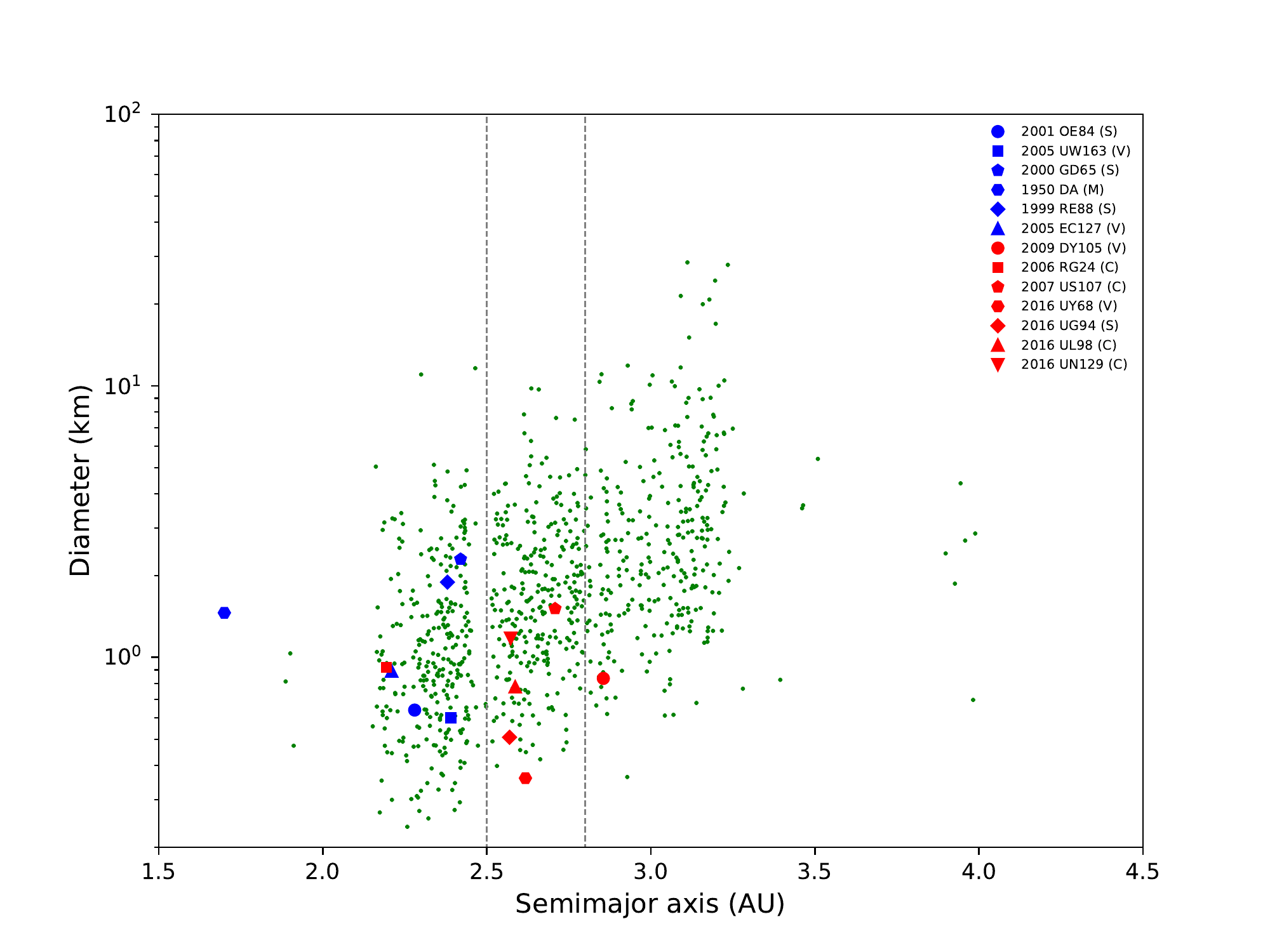}
\caption{Asteroid diameter vs.\ semi-major axis. The green filled circles are asteroids with reliable rotation periods (green) and the gray ones are asteroids with 10 or more detections. The dashed lines show the divisions of empirical geometric albedo in $w_{P1}$ band used for diameter calculations for asteroids in different locations. The blue and red symbols are the six known SFRs and the seven PS1-SFRs, respectively. Their designations are given on the plot.}
\label{a_d}
\end{figure}

\begin{figure}
\plotone{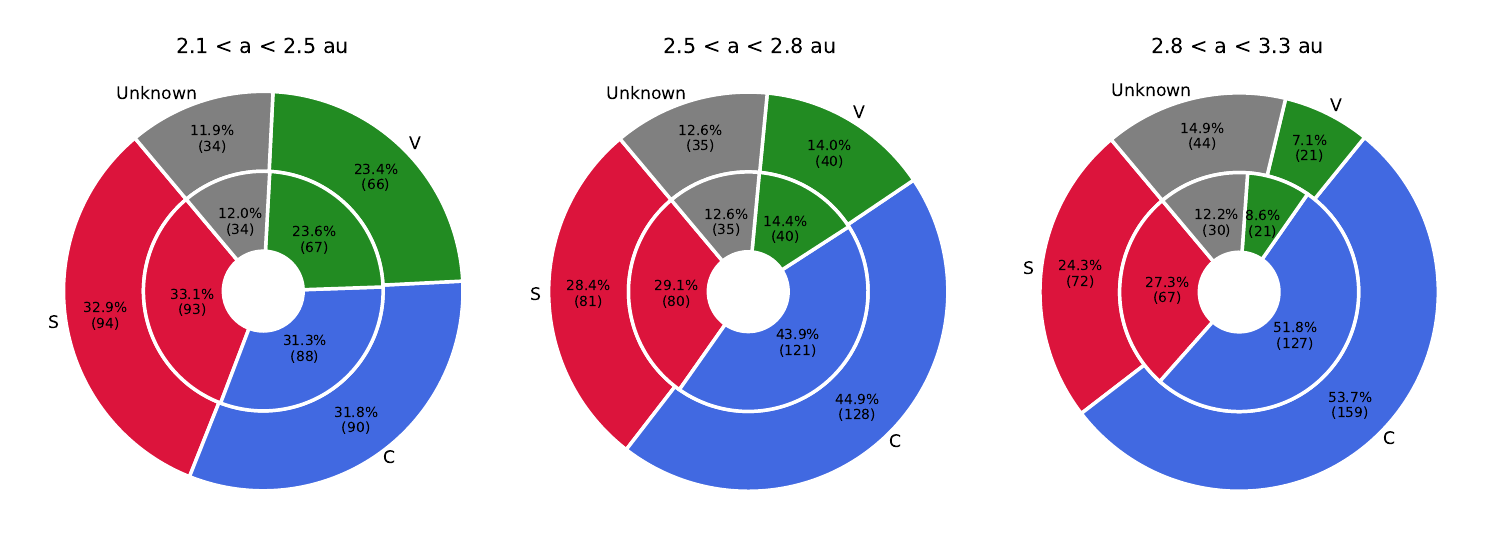}
\caption{The spectral type distributions of our samples with qualified color measurements in the inner (left), mid (middle), and outer (right) main belt. The outer annulus includes all sizes and the inner one is limited to the diameters of $D < 5$ km. The percentage and the number in each bin are give on the plot.}
\label{sp_dist}
\end{figure}

From our survey, 3858 asteroid light curves with 10 detections or more in $w_{P1}$ band were extracted, in which 876 have reliable measurements for their rotation periods (i.e., $U >= 2$). Their magnitude distribution is shown in Fig.~\ref{mag_hist}, where we see the recovery rate of rotation period is decreasing toward the faint end. Most of our samples are MBAs, and the rest includes some Hungaras, Cybeles, and Hildas. The diameter range of our samples can be found in Fig~\ref{a_d} which shows the plot of their semi-major axes vs diameters. Among the 876 asteroids with reliable rotation periods, 762 have qualified color measurements for spectral type determinations. Their spectral distributions were divided by the inner ($2.1 < a < 2.5$ au), mid ($2.5 < a < 2.8$ au), and outer ($a > 2.8$ AU) main belt and shown in Fig.~\ref{sp_dist}. We see that the C-type becomes more dominant with greater heliocentric distance. The detailed information of 876 asteroids with reliable rotation periods are listed in Table~\ref{table_p}, and their folded light curves are shown in Figs.~\ref{lightcurve00}-\ref{lightcurve17}.

\begin{figure}
\plotone{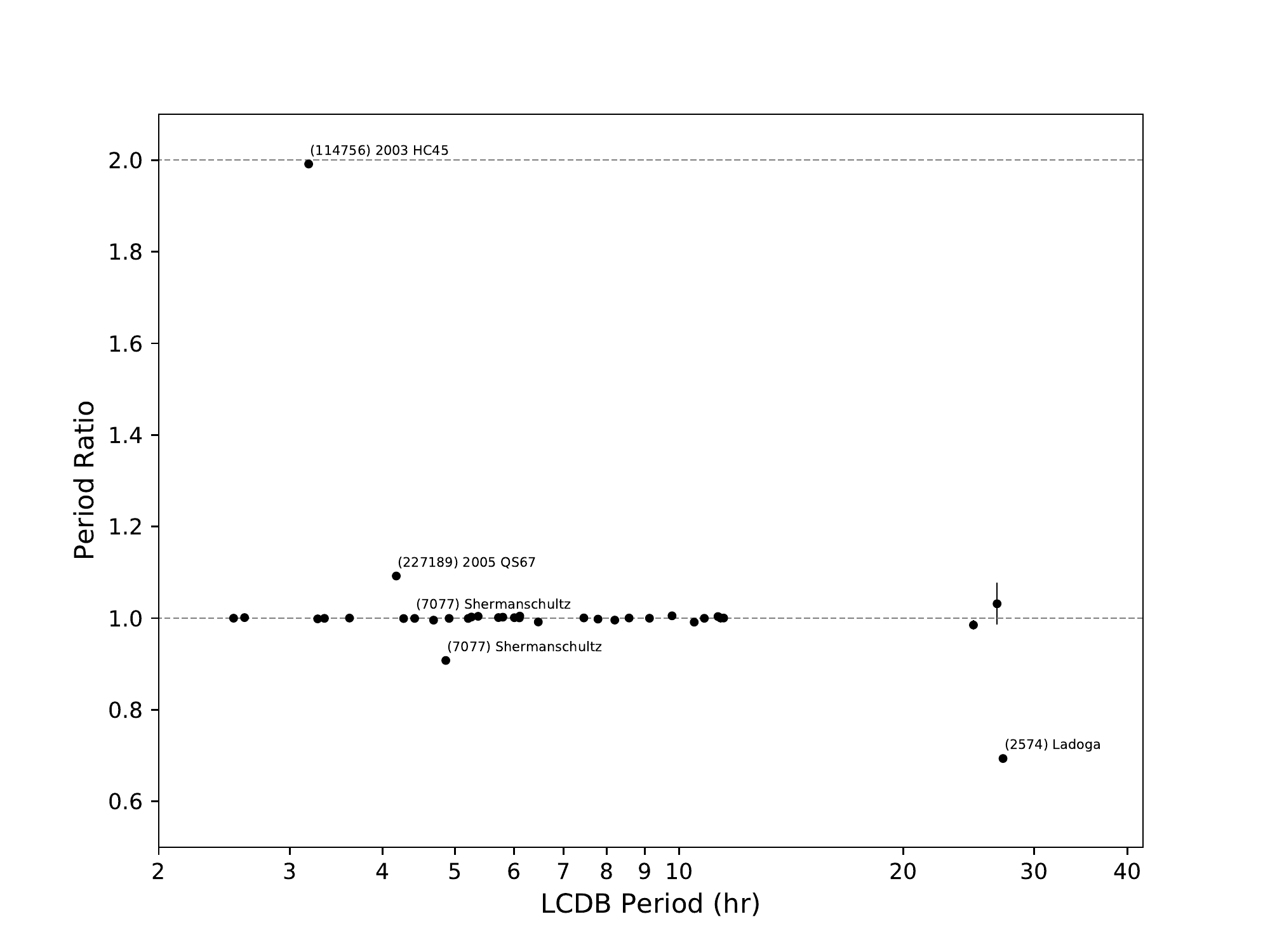}
\caption{Rotation period comparison for asteroids both available in this work and the LCDB. The rotation-period ratios of of 34 asteroids are shown in this plot. Note that only $U \ge 2$ objects in the LCDB are used in the comparison.}
\label{comp_period}
\end{figure}

Among the 876 asteroids with reliable rotation periods, 34 of them also have a rotation period of $U \ge 2$ listed in the LCDB\footnote{The light-curve database \citep{Warner2009}; http://www.minorplanet.info/lightcurvedatabase.html.}. Therefore, we compare their rotation periods in both data sets. The ratios of rotation period from our survey to the LCDB are shown in Fig.~\ref{comp_period}, where we see that most objects have consistent results except for four objects showing difference greater than 5\%. Because our observation time-span was only a few days, it was difficult to recover long rotation period. If possible, we mostly have a folded light curve with partial coverage of a full rotation, like asteroid (2574) Ladoga in Fig.~\ref{comp_period}. Therefore, this kind of long rotation period obtained from our survey can be seen as a lower limit for these objects. The other three cases are briefly discussed below. For asteroid (114756) 2003 HC45, we derived a rotation period of 6.33 hour, which doubles the value given in \citet{Chang2015}. While our folded light curve of 2003 HC45 was assigned as $U = 3$ for its significant double-peak feature, that of \citet{Chang2015} was assigned as $U = 2$ and only shows a single-peak feature with a insignificant secondary dip. Therefore, we believe that \citet{Chang2015} identified a half of the actual rotation period for 2003 HC45. For asteroid (7077) Shermanschultz, \citet{Waszczak2015} published two rotation periods, 4.41 and 4.86 hours, using 29 and 28 data points, respectively. Comparing to this, our result, 4.41 hour, from a folded light curve of $U = 3$ with much more data points densely covering in the rotational phase, is consistent with the former. Therefore, we believe that we have high reliability on the rotation period of (7077) Shermanschultz and the 4.41 hour is the actual value. For asteroid (227189) 2005 QS67, we derived a rotation period of 4.55 hours, which is close to 4.17 hour given by \citet{Chang2015}. Both folded light curves were assigned as $U = 2$ and look equally good. Therefore, its rotation period needs further confirmation. Since the difference is less than 10\%, we therefore see this case as a consistent result. In general, our rotation period measurements are reliable for the following analysis.

\subsection{The 2-hour Spin-Rate Limit}
The 2-hour spin-rate limit shown in the asteroids of $D > 150$ m has been seen as a supporting evidence for the rubble-pile structure \citep{Pravec2002}. Although the six large SFRs show contradictory to the concept of rubble-pile structure, the chance to discover a large SFR is still very rare \citep[see Table 2 in][ and the references therein]{Chang2017}. This is also the case in our survey that only seven out of the 876 reliable rotation periods were found to be shorter than 2 hours (detailed analysis please see below). Fig.~\ref{dia_per} shows the plot of diameters vs. rotation periods of our samples, where we see an obvious stop around 2 hours. Although the chance to find an object with rotation period shorter than 2 hour is higher in our survey (i.e., $\sim 1$ \%) than \citet[][i.e., $\sim 0.1$ \%]{Chang2015, Chang2016}, the rubble-pile structure is still a reasonable explanation to what we observed.

\subsection{The Large Super-Fast Rotators}
In our survey, eight objects were found to have reliable rotation periods of $< 2$ hours. Their period analysis was given in Fig.~\ref{ps1_sfr_lc}, in which all the rotation periods are clearly detected on the periodograms and all the folded light curves show a clean trend. \citet{Harris2014} pointed out that small light-curve amplitude (i.e., less than 0.2-0.3 mag) can possibly be dominated by the 4th or 6th harmonics that leads to a detection of a half or one-third of the actual rotation period. To test this possibility, we used the 4th-order Fourier series to run the analysis for these eight objects again. Figure~\ref{ps1_sfr_lc_4th} shows the periodograms and the folded light curves of the 4th-order Fourier series fitting, where we see all the fittings have been improved in someway due to the better fitting in detailed features. The best-fitted periods of 4th-order fitting are consistent with the previous 2nd-order fitting except for 2001 FQ10 and 2016 UL98 that their best fitted periods of 4th-order analysis are double the periods of the pervious 2nd-order fitting. For 2001 FQ10, its 4th-order folded light curve enhances a very insignificant 3rd peak that was missed in the previous 2nd order fitting and gives 3.38 hours as the best-fitted period, and therefore, we exclude this objects as a SFR for now and wait for further confirmation of its rotation period. For 2016 UL98, the folded light curve of the 4th-order fitting shows a very insignificant difference in the depths of the 1st deep and the 3rd deep. However, we doubt this difference is due to the scattered data points. If the difference is true, the new period (i.e., 1.04 hour) is still shorter than 2 hours and this reminds 2016 UL98 as a SFR as well. Therefore, we use 0.52 hour as the rotation period of 2016 UL98 in the following discussion. The detailed information of these seven objects (hereafter, PS1-SFRs) along with the previous reported large SFRs can be found in Table~\ref{sfr_tbl}.


The diameter range of the PS1-SFRs is from $\sim 0.3$ to $\sim 1.5$ km\footnote{The diameters of the PS1-SFRs are estimated based upon the assuming albedos of their spectral types. For the neutral colored objects (i.e., SDSS $a^* < 0$), the diameter would be reduced by a factor of two when assuming E-type \citep[i.e., 0.45;][]{Demeo2013} instead of C-type. However, this still gives the diameter estimations of four neutral colored PS1-SFRs of $\gtrsim 0.3$ km. The details of spectral types of the PS1-SFRs can be found in below.}. Using $P \sim 3.3 \sqrt{(1 + \Delta m)/\rho}$, the minimal bulk density to maintain the equilibrium between self-gravity and centrifugal force for a rubble-pile asteroid can be calculated \citep{Harris1996}. Fig.~\ref{spin_amp} shows the plot of the spin rates vs. light-curve amplitudes of our samples with the spin-rate limits calculated for bulk densities of $\rho = 3, 4$ and 5 g cm$^{-3}$, where we see the PS1-SFRs all need a relatively high bulk density to survive under their super-fast rotation. In addition to the PS1-SFRs, another asteroid of $D \sim 0.7$ km, 2016 UK50, also requires a bulk density of $\rho > 4$ g cm$^-3$ to keep intact, although its rotation period is only 2.2 hours. Such high bulk density is unusual among asteroids \citep[see Table 2 in][]{Demeo2013}. Therefore, the PS1-SFRs and 2016 UK50 are very unlikely to be explained simply by the rubble-pile structure. Is it possible that these PS1-SFRs are large monoliths? Although we have no evidence to totally rule out this possibility, the question becomes how they could avoid numerous collisions or keep these numerous impacts not destructive \footnote{Given the intrinsic collision probability of main belt asteroids shown by \citet{Polishook2016} as $N_{impacts} = P_i N(> r_{projectile} (r_{target} + r_{projectile})^2)$, where $P_i = 2.85 \times 10^{-18}$ km$^{-2}$yr$^{-1}$ and $r_{projectile}$ is 16 meter \citep{Bottke1994}, the PS1-SFRs would have $10^3 - 10^4$ collisions during 1 Gyr.}.

The color calculations of the PS1-SFRs are shown in Fig.~\ref{ps1_sfr_color}. Except for 2016 UL98, the color measurements of the other six PS1-SFRs all have good agreement with the folded light curve in $w_{P1}$ band. Although the color measurement of 2016 UL98 are relatively scattered, they are still within their light-curve variation in $w_{P1}$ band. In addition, 2016 UN129 might also have great uncertainty in its color measurements because it was relatively faint and had only one detection in each $g_{P1}$, $r_{P1}$, and $i_{P1}$ bands. Fig.~\ref{ps1_sfr_sdss} shows the plot of the SDSS $a^*$ vs $i-z$ for the seven PS1-SFRs on top of the objects with meaningful color calculation in our survey. Note that we adopt the photometric error for the color uncertainty. As seen, most of our samples populate in the dense region of SDSS sampled asteroids \citep[see Fig. 3 in][]{Parker2008}, and the seven PS1-SFRs have diversity in their colors. Among them, 2016 UN129 has an unusual location on the plot that might be due to its relatively large uncertainty in its colors measurements. Using the boundary defined in \citet{Parker2008}, the colors of the seven PS1-SFRs suggest that 2016 UG94 is S-type, 2009 DY105 and 2016 UY68 are V-type, and the other four are C-type\footnote{Considering most E-type objects are found in Hungarias and the population of M-type objects are relatively small in mid main belt \citep{Demeo2013}, we believed these four neutral colored SFRs are very likely to be C-type.}. \citet{Chang2017} pointed out a possible taxonomic tendency in the six known large SFRs that none of them are C-type asteroids. However, the diverse colors of the seven PS1-SFRs seem to rule out that tendency. Although the spectral types of the seven PS1-SFRs need further confirmations, our result suggests that large SFRs in the main belt can have different compositional materials.

Using the Drucker-Prager yield criterion, \citet{Holsapple2007} showed that SFRs can survive with the presence of internal cohesion. Following the equations and calculations shown in \citet{Holsapple2007, Rozitis2014, Polishook2016, Chang2017}, we estimated the cohesion needed for the PS1-SFRs assuming bulk density $\rho = $ 2.72, 1.93, and 1.33 for S-, V-, and C-type, respectively \citep{Demeo2013}. The smallest cohesion of the PS1-SFRs is $\sim10~Pa$ of 2016 UY68 and the largest is $\sim700~Pa$ of 2016 UL98. This cohesion range is is similar to that of the known large SFRs (see Table~\ref{sfr_tbl}) and the lunar regolith \citep[i.e., 100 to 1000 $Pa$;][]{Mitchell1974}. This probably suggests similar source of generating cohesion for these large SFRs.

Unlike the six known large SFRs which belong to either near-Earth objects or inner MBAs, the PS1-SFRs populate throughout the main belt. This suggests that large SFRs can form in any location of the main belt. However, it is very interesting to notice that six out of the seven PS1-SFRs locate in the mid main belt. If large SFRs are uniformly distributed in the main belt and have similar sizes (i.e., about 1 km), a general survey for asteroid rotation period, like ours, should have more chance to discover them in the inner main belt (i.e., better photometric accuracy for asteroids with the same size). As shown by the simulation of deriving rotation period, the chances to recover spin rate of $f >= 3$ rev/day are very similar at a fixed magnitude and a fixed amplitude (see Fig.~\ref{debias_map}). Therefore, less large SFRs being detected in the inner main belt is not because we miss to derive their rotation periods. Do we obtain more reliable rotation periods in the mid main belt to detect more large SFRs there? When limiting the diameter range to $0.3 - 2$ km, we have 237 and 193 reliable rotation periods in the inner and mid main belt, respectively. Therefore, this is not the case for our survey. A possible explanation is that less SFRs exist in the inner main belt. While the detection rate of large SFRs in our survey is only $\sim0.4\%$ (i.e., 1 out of 237 reliable rotation periods) in the inner main belt for objects of $0.3 < D <2$ km, the mid main belt is $\sim3.1\%$ (i.e., 6 out of 193). This can also explain why the chance to discover SFRs was lower (i.e., $\sim 1$ out of 1000) in the pervious similar surveys \citep[e.g.,][]{Chang2014a, Chang2015, Chang2016} than this work (i.e., $\sim 1$ out of 100). Because the previous surveys were merely able to detect kilometer-sized asteroids in the mid main belt.


\subsection{The Spin-Rate Distributions}
We first carried out the spin-rate distributions according to their sizes and locations in the main belt. The samples were divided into inner, mid, and outer MBAs with diameters of $3 < D < 15$ km and $D < 3$ km. Moreover, we followed the approach shown by \citet{Masiero2009} and \citet{Chang2015} to consider the possible observational biases in our survey. Fig.~\ref{debias_map} is the recovery rates of rotation period for different magnitudes in the simulation of our survey. In general, it tends to have higher recovery rate for brighter, short-period, and large-amplitude objects. The de-biased results are given in Fig.~\ref{spin_rate_comp}. Because we only have a small number of asteroids of $3 < D < 15$ km in the inner and mid main belt, and we therefore exclude them in the following discussion. Overall, our results are very similar to that of \citet{Chang2015}.

For asteroids of $3 < D < 15$ km in the outer main belt, we see the spin-rate distribution showing a smooth decline in number along the spin-rate. This means that the asteroid system is not in collisional equilibrium, otherwise it would have a Maxwellian spin-rate distribution \citep{Salo1987}. Although the YORP effect can deviate the distribution from a Maxwellian form, we not clear how the one like ours can be produced.

For asteroids of $D < 3$ km, a significant drop in number is observed at the spin-rate of $f = 5$ rev/day in all locations. As pointed out by \citet{Chang2015}, the high spin-rate bins only contain very few small and elongated objects.  This is also the case for our survey, in which most fast rotators of $D < 3$ km also have small amplitudes (i.e., $< 0.4$ mag; see the green line in Fig.~\ref{spin_rate_comp}). \citep{Chang2015} suspected that the rotational disruption generates the deficiency in small and elongated fast rotators. Because the spin-rate limit for small and elongated objects is lower and their YORP timescales, moreover, are also shorter than large objects, these objects could have been pushed through the spin-rate limit and destroyed already.

Therefore, a comprehensive simulation on the spin-rate evolution for the entire main asteroid belt is needed to understand what we see here.

\section{Summary}
Using the PS1, we conducted a survey for asteroid rotation period during October 26-31, 2016, from which more than 1500 new asteroids were reported to the Minor Planet Center, 3858 asteroid light curves with 10 or more detections were extracted, and 876 reliable rotation periods were obtained. The spin-rate distributions for asteroids of different sizes and locations in the main belt are similar to \citep{Chang2015, Chang2016}, which shows (a) the number of asteroid decreases along with spin rate for asteroids of $D > 3$ km; (b) a number drop appears at $f = 5$ rev/day for asteroids of $D < 3$ km; and (c) no obvious dependence on the location was found.

Among the 876 reliable rotation periods, only seven objects were found to have rotation periods shorter than 2 hours. This suggests that SFRs are still rare. Considering the significant difference in number between SFRs and the rest in our survey, it looks like the rubble-pile structure still can explain our observation.

Assuming a rubble-pile structure, the seven PS1-SFRs require relatively high bulk density to keep intact under their super-fast rotation. Such high bulk density is unusual to asteroids, and, we therefore believe other physical strengthes, in addition to self-gravity, are needed to explain them. Using the Drucker-Prager yield criterion, the cohesion for the PS1-SFRs were estimated in a range of $\sim 10 - 600 Pa$, which is similar to that of the six known large SFRs and the lunar regolith \citep{Mitchell1974}. This might suggest that SFRs could share similar source to generate internal cohesion. Unlike the six known large SFRs locating in inner main belt or near-Earth region, the PS1-SFRs populate throughout the main asteroid belt. Moreover, the diverse colors of the seven PS1-SFRs rule out the possible taxonomic tendency previously found in the six known large SFRs. This suggests that the formation of SFR is unlikely to have dependence on location and composition. However, it is interesting that five out of the the seven PS1-SFRs are mid MBAs. Considering the survey condition, we suspect that mid main belt possibly harbors more SFRs than the inner main belt.

\acknowledgments This work is supported in part by the National Science Council of Taiwan under the grants MOST 107-2112-M-008-009-MY2, MOST 104-2112-M-008-014-MY3, MOST 104-2119-M-008-024, and MOST 105-2112-M-008-002-MY3, and also by Macau Science and Technology Fund No. 017/2014/A1 of MSAR. We thank the referee, Dr. Alan Harris, for his useful comments and suggestions to improve the content of this paper.

\clearpage

\clearpage

\begin{figure}
\plotone{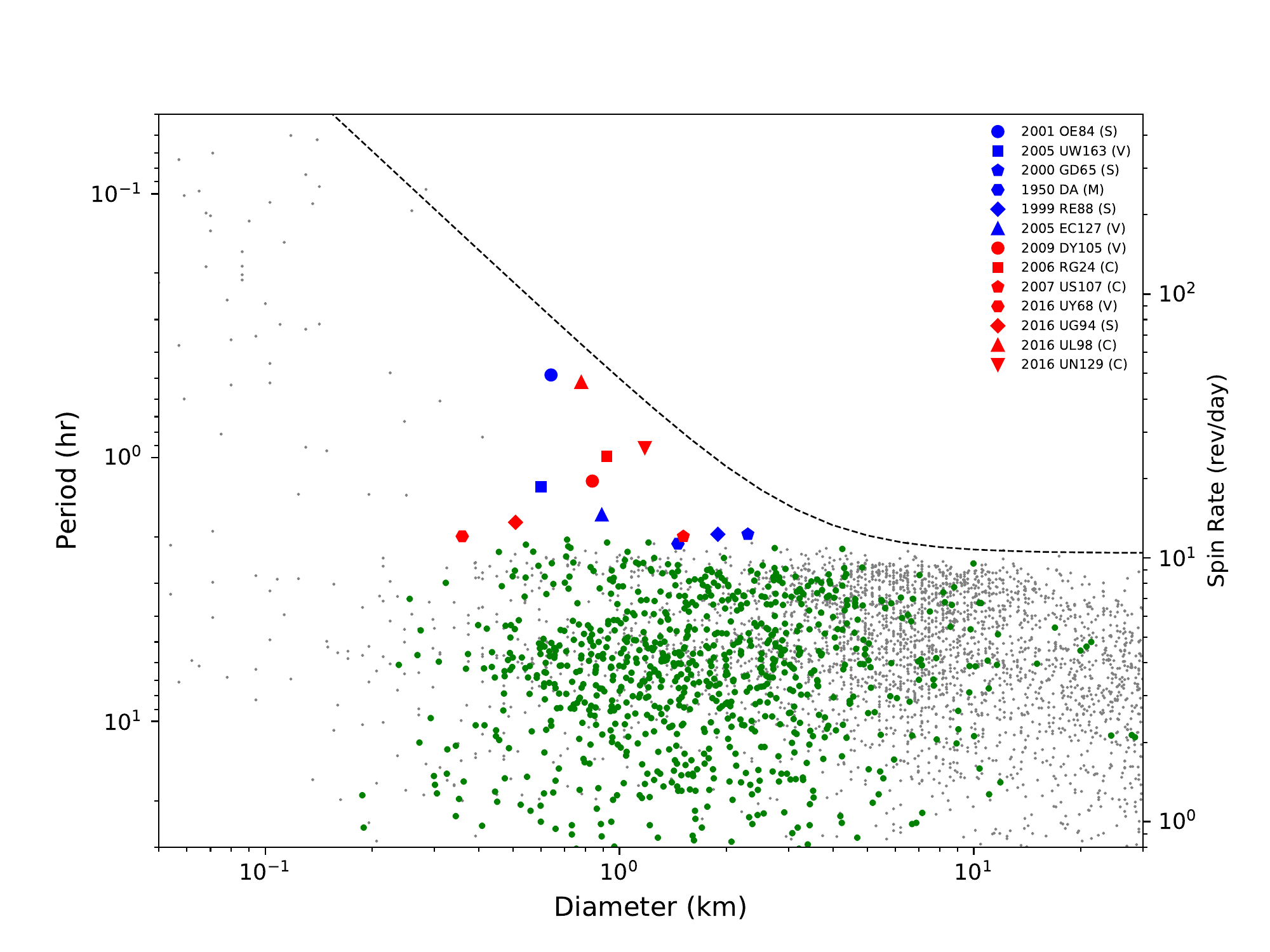}
\caption{Asteroid rotation period vs.\ diameter. The green and gray filled circles are the asteroids with reliable rotation periods in this work and the LCDB objects of $U \geq 2$, respectively. The six known SFRs are shown with blue symbols and the seven PS1-SFRs are in red. The dashed line is the spin-rate limit with internal cohesion adopted from \citet{Holsapple2007}.}
\label{dia_per}
\end{figure}

\begin{figure}
\plotone{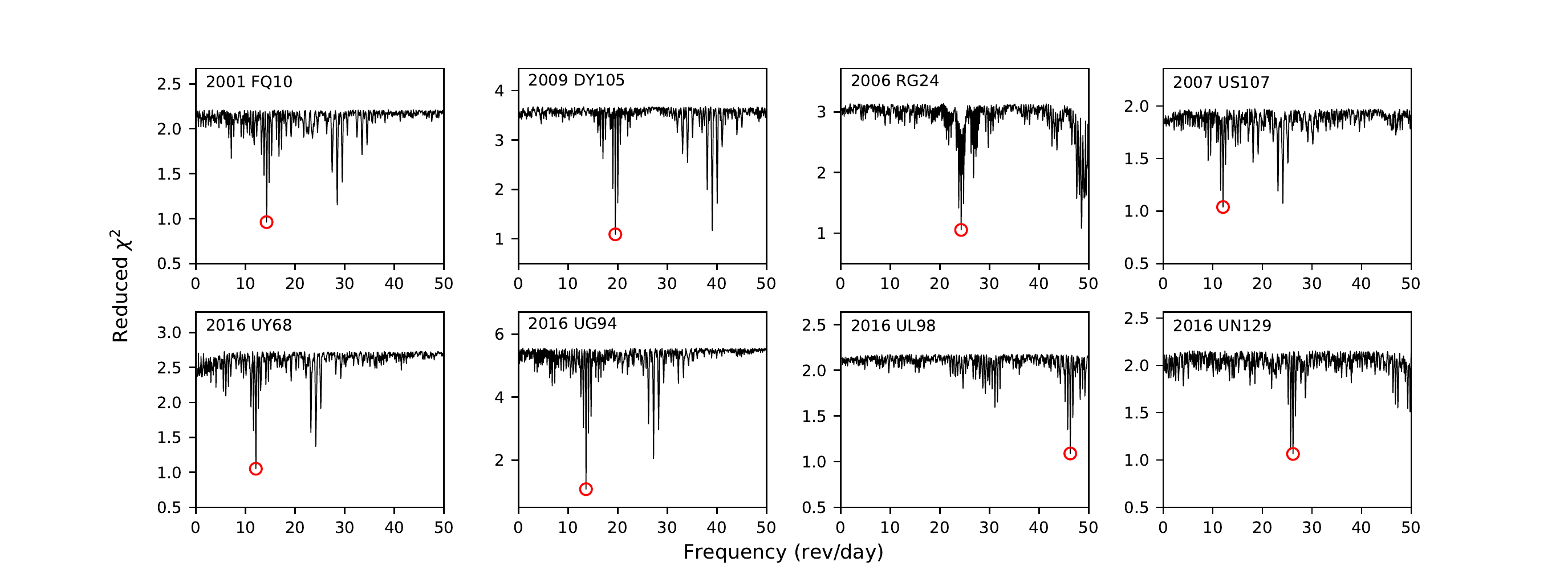}
\plotone{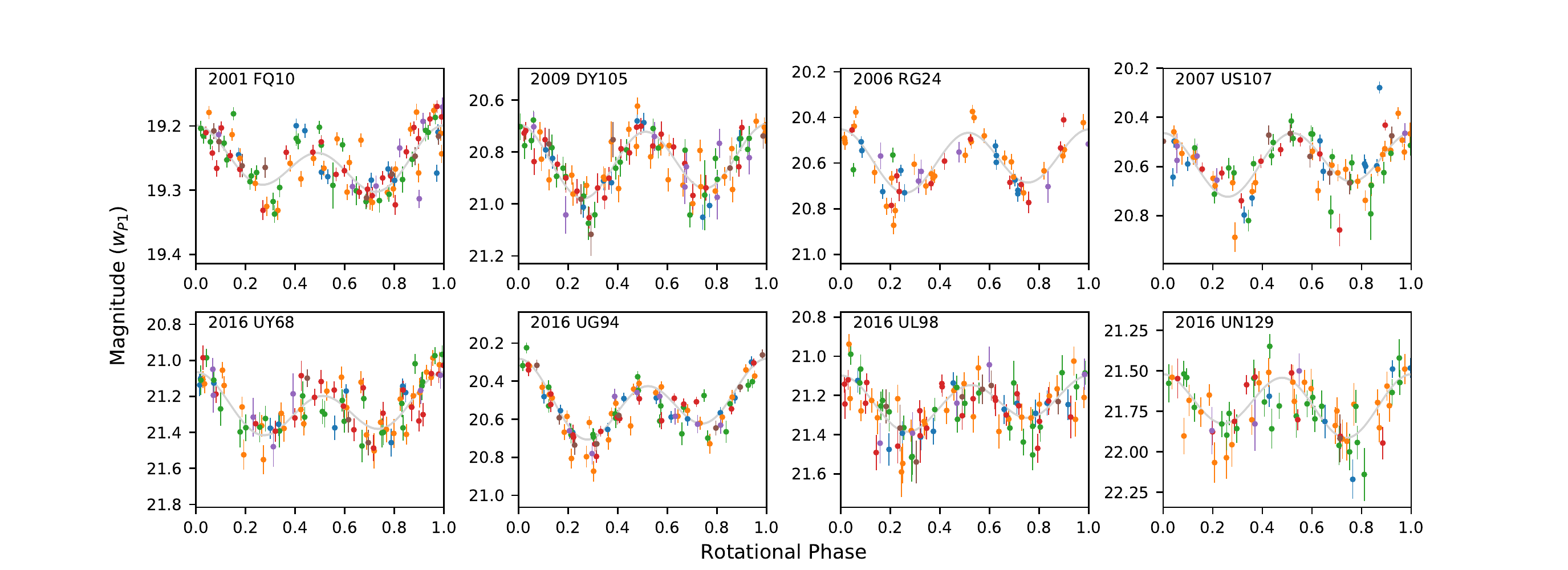}
\caption{The period analysis for the eight objects of $P < 2$ hours using 2nd-order Fourier series. Upper: periodogram; lower: folded light curve, where colors means data points taken in different nights.}
\label{ps1_sfr_lc}
\end{figure}

\begin{figure}
\plotone{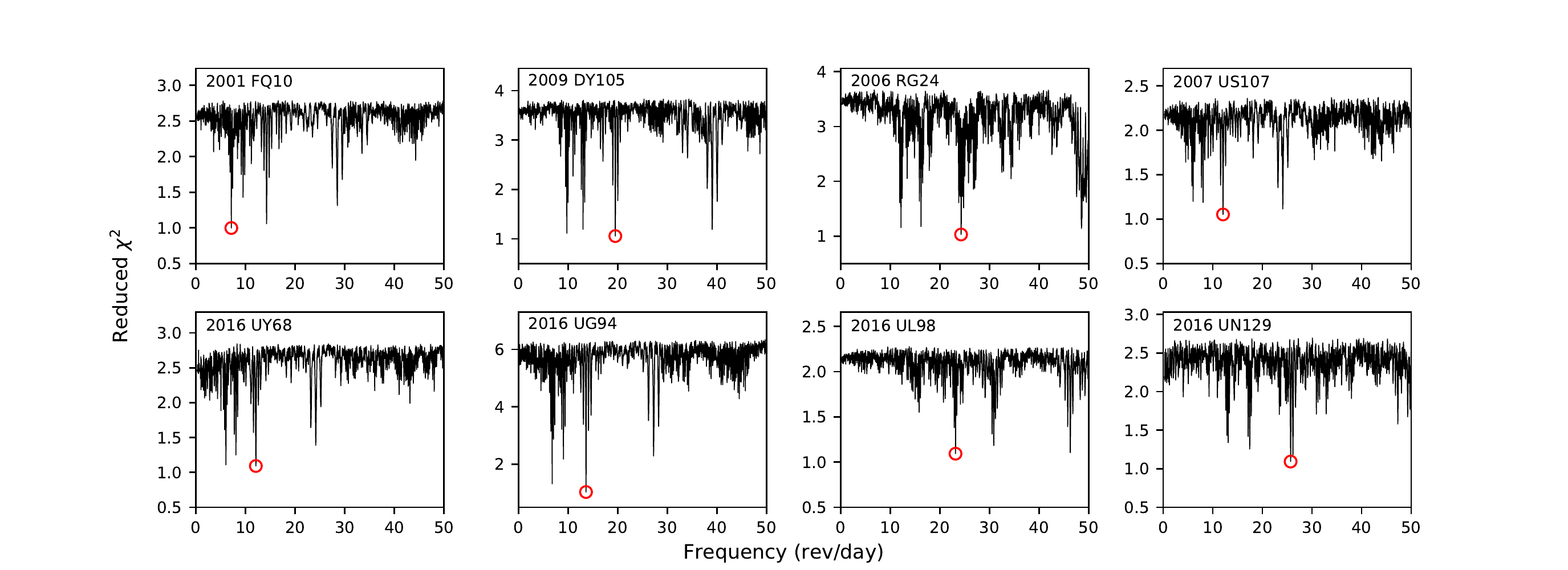}
\plotone{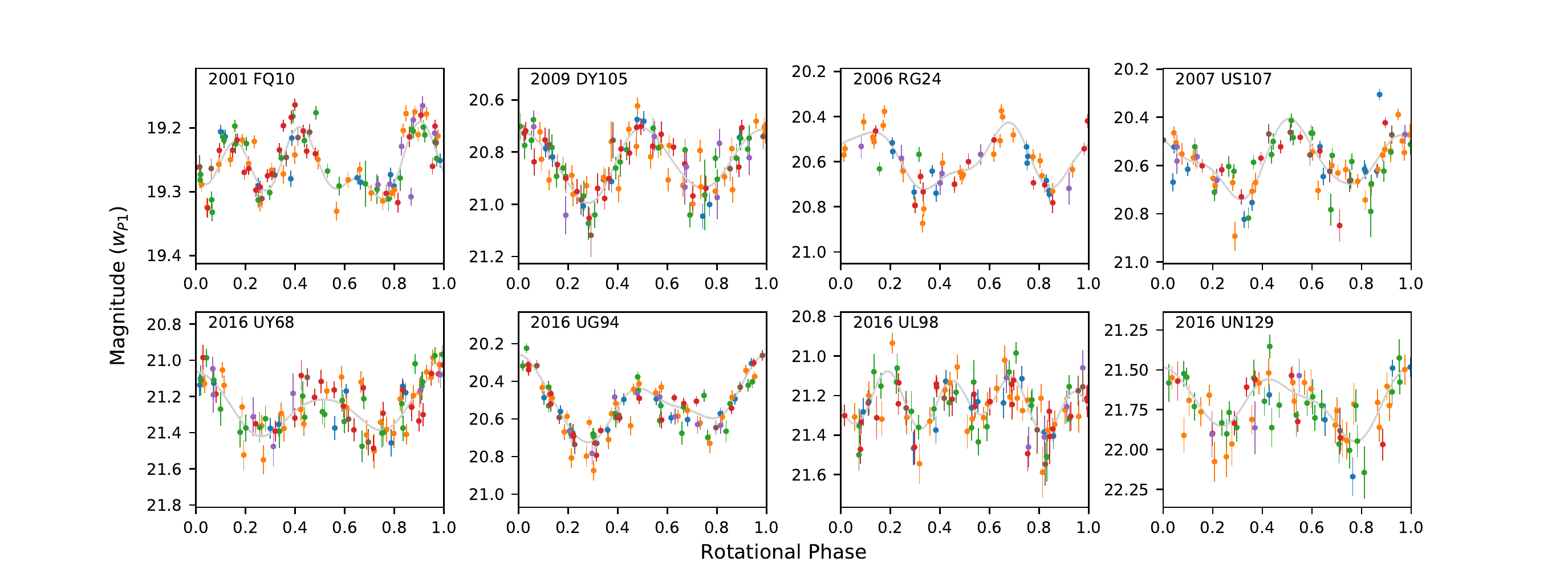}
\caption{The period analysis for the eight objects of $P < 2$ hours using 4th-order Fourier series. Upper: periodogram; lower: folded light curve, where colors means data points taken in different nights.}
\label{ps1_sfr_lc}
\end{figure}


\clearpage
\begin{figure}
\plotone{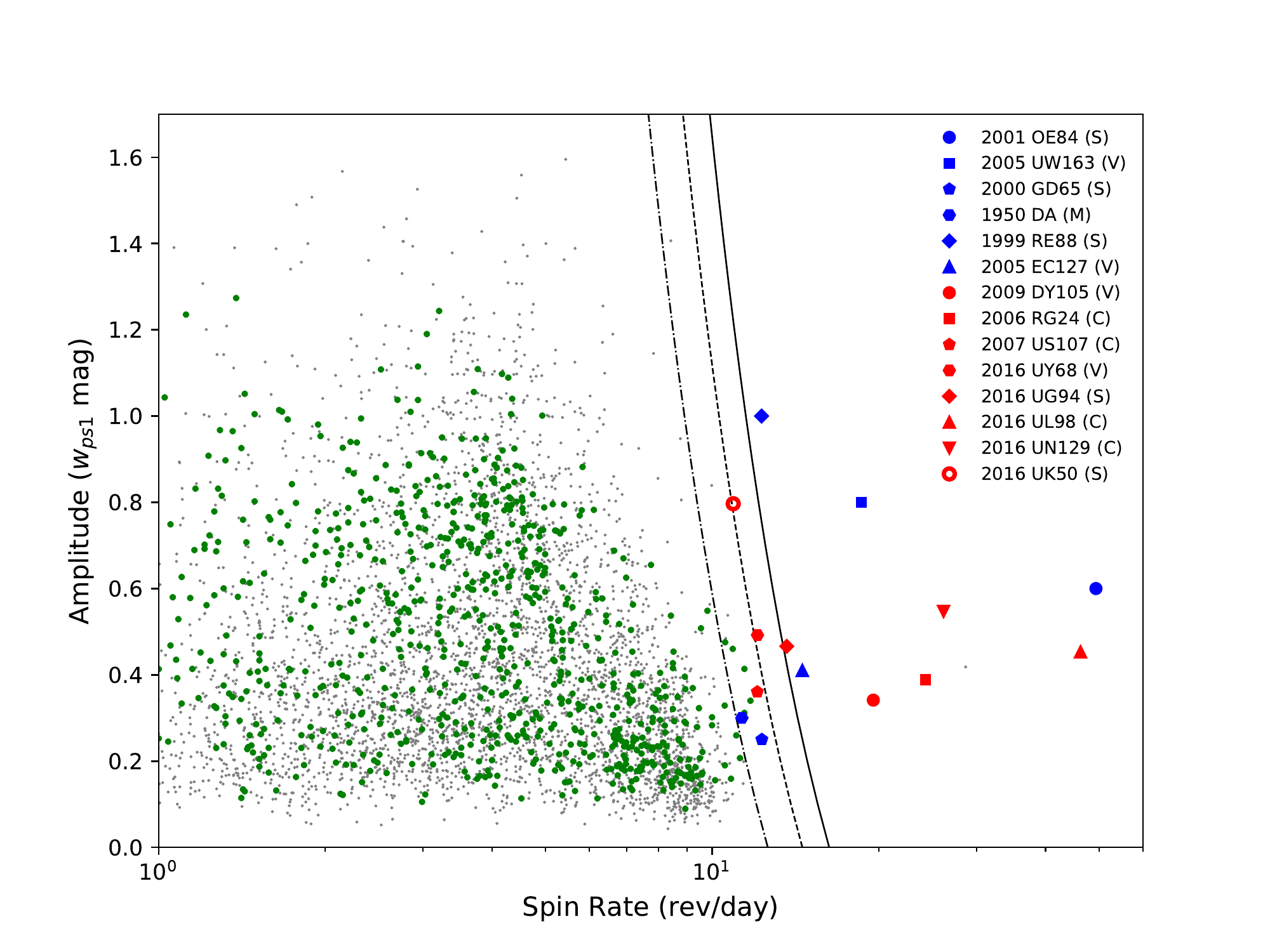}
\caption{Lightcurve amplitude vs.\ spin rate. The symbols are the same with Fig.~\ref{dia_per}. The dashed, dot-dashed and dotted lines represent the spin-rate limits for rubble-pile asteroids with bulk densities of $\rho =$ 5, 4, and 3~g/cm$^3$, respectively, according to $P \sim 3.3 \sqrt{(1 + \Delta m)/\rho}$ \citep{Pravec2000}. Note that the asteroids of $D < 0.3$ km are not included in this plot.}
\label{spin_amp}
\end{figure}

\begin{figure}
\plotone{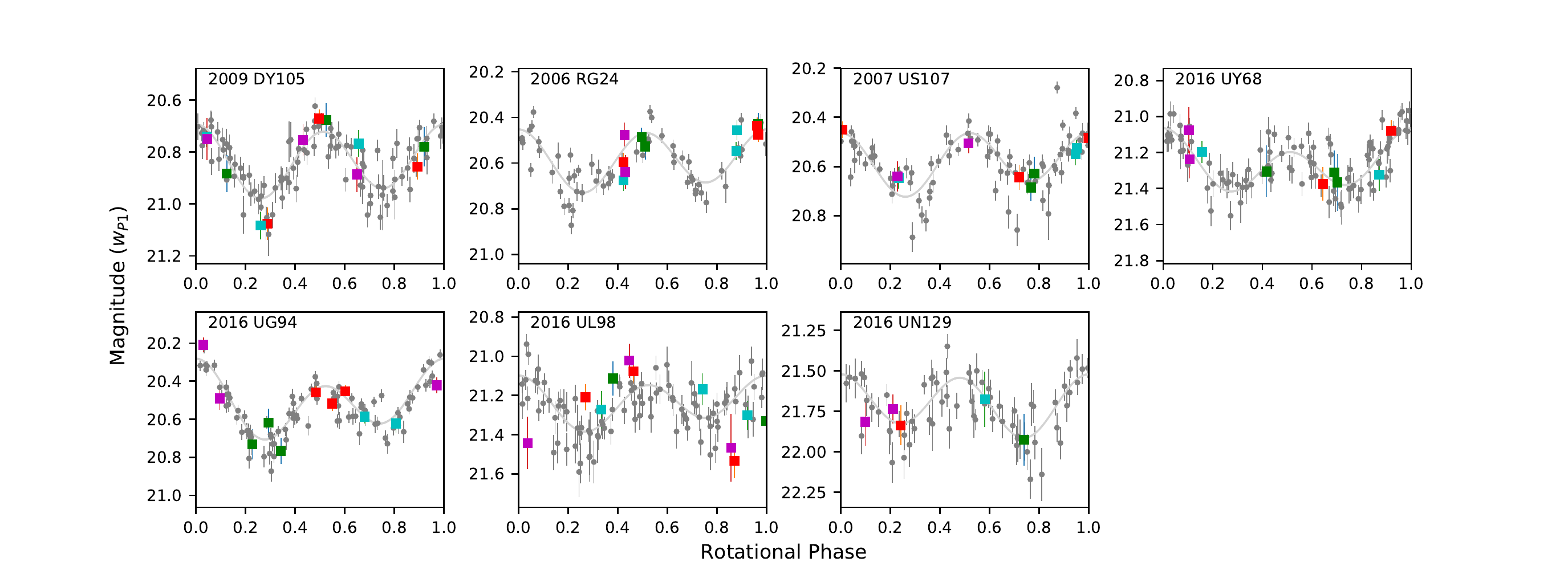}
\caption{The color calculation for the PS1-SFRs. The gray filled circles are data points in $w_{P1}$ band. The green, red, canyon and magenta filled squared are the data points of $g_{P1}$, $r_{P1}$, $i_{P1}$, and $z_{P1}$, respectively, and their magnitudes are shifted to match the $w_{P1}$-band folded light curve.}
\label{ps1_sfr_color}
\end{figure}

\begin{figure}
\plotone{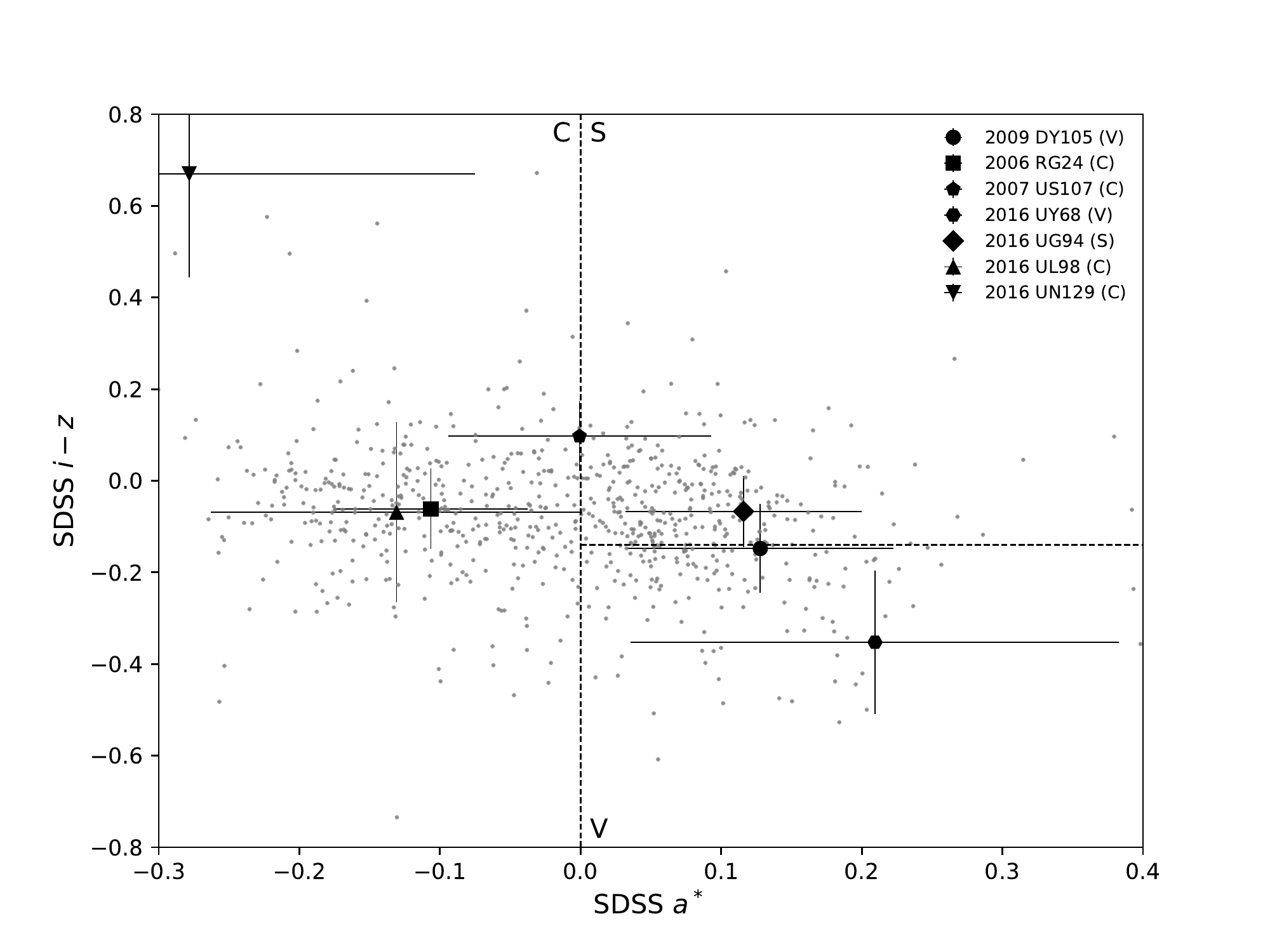}
\caption{The plot of $a^*$ vs.\ $i-z$ in SDSS color for the asteroids with reliable rotation periods and meaningful color calculations. The seven PS1-SFRs are indicated with black symbols with the designation given on the plot. Note that the boundary for S-, C-, and V-type asteroids are adopted from \citet{Parker2008}.}
\label{ps1_sfr_sdss}
\end{figure}

\begin{figure}
\plotone{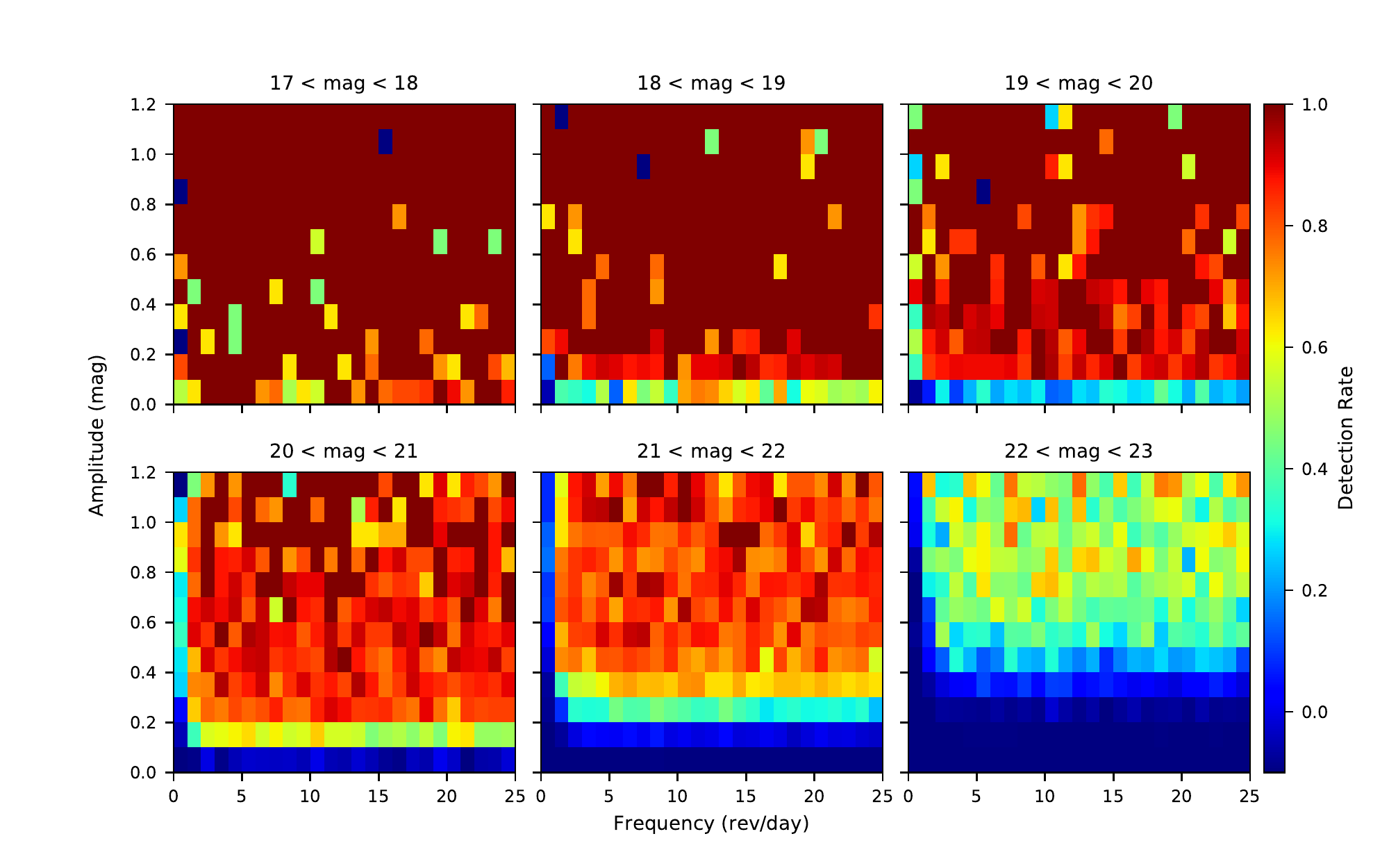}
\caption{Recovery rate for asteroid rotation period. The color bar represents the recovery rate. The apparent magnitude intervals are given on the top of each plot.}
\label{debias_map}
\end{figure}

\begin{figure}
\plotone{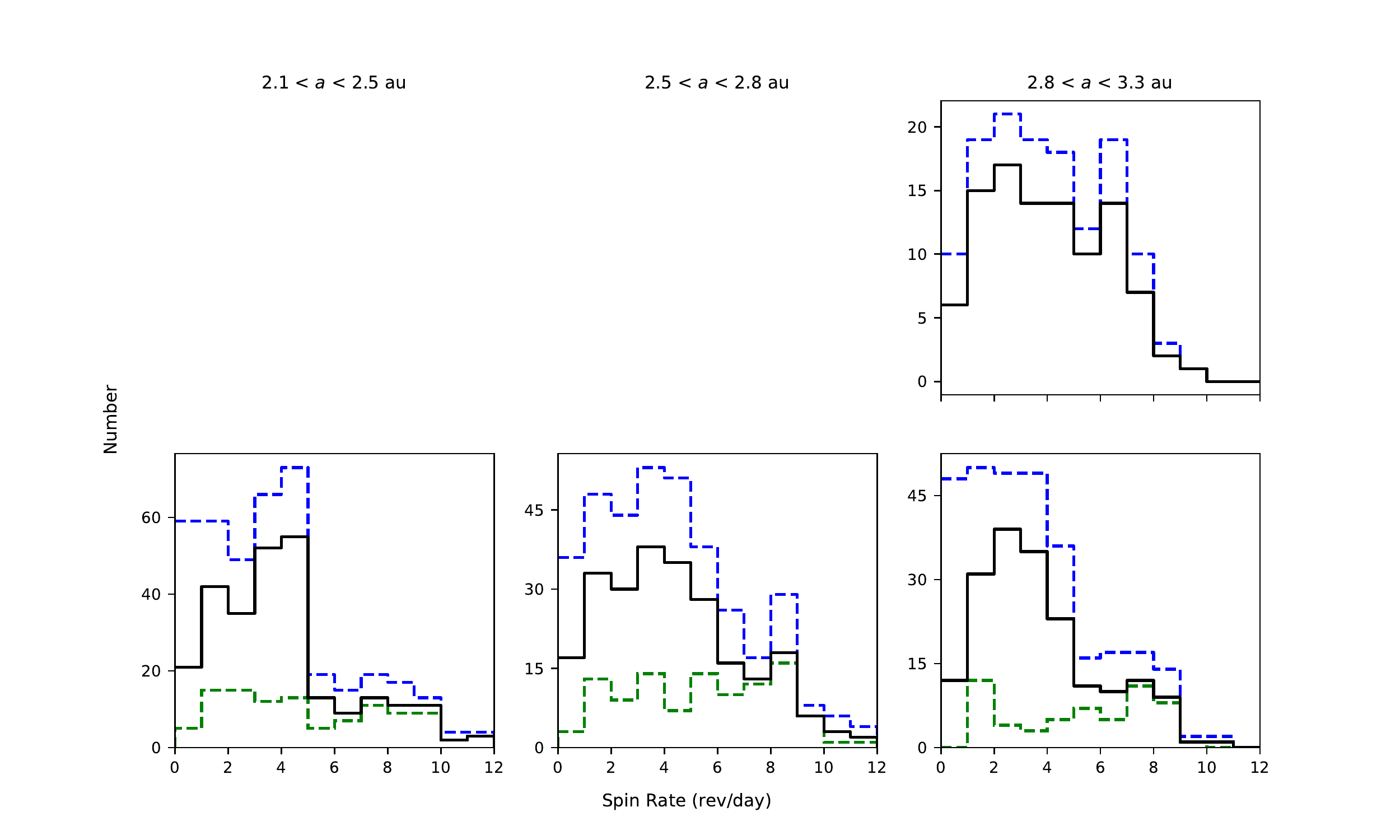}
\caption{The spin-rate distributions for asteroids with diameter of $3 < D < 15$ km (top) and $D < 3$ km (bottom) at inner (left), mid (middle) and outer (right) main belt. The black line and blue dashed line are the original distribution and that after de-bias, respectively. The green dashed line is the objects with amplitudes of $\Delta m < 0.4$ mag. Note that the numbers of asteroids with diameter of $3 < D < 15$ km in the inner and mid main belt are too small to carry out a meaningful statistics.}
\label{spin_rate_comp}
\end{figure}

\clearpage
\begin{deluxetable}{cccccccc}
\tabletypesize{\scriptsize}
\tablecaption{Observation information. \label{obs_log}}
\tablewidth{0pt}
\startdata \tableline\tableline
Field No. & [RA,  Dec.] & Oct 26 2016 & Oct 27 2016 & Oct 28 2014 & Oct 29 2014 & Oct 30 2014 & Oct 31 2014\\
         & [$^{\circ}$,  $^{\circ}$] & $\Delta$t, N$_\textrm{exp}$ & $\Delta$t, N$_\textrm{exp}$ & $\Delta$t, N$_\textrm{exp}$ & $\Delta$t, N$_\textrm{exp}$ & $\Delta$t, N$_\textrm{exp}$ & $\Delta$t, N$_\textrm{exp}$ \\
\tableline
     1 &  [28.10, 13.14] &  6.8, 12 &  6.6, 32 &  6.5, 31  &  5.7, 25 &  6.9, 8 &  1.7, 8 \\
     2 &  [29.17, 10.33] &  6.3, 11 &  6.6, 32 &  6.5, 31  &  5.7, 25 &  6.9, 8 &  1.7, 8 \\
     3 &  [31.00, 14.18] &  6.8, 12 &  6.6, 32 &  6.5, 31  &  5.7, 25 &  6.9, 8 &  1.7, 7 \\
     4 &  [32.04, 11.36] &  6.8, 12 &  6.6, 32 &  6.5, 31  &  5.7, 25 &  6.9, 8 &  1.7, 8 \\
     5 &  [33.92, 15.18] &  6.8, 12 &  6.6, 32 &  6.5, 30  &  5.7, 25 &  6.9, 8 &  1.7, 7 \\
     6 &  [34.93, 12.35] &  6.3, 11 &  6.6, 32 &  6.5, 31  &  5.7, 25 &  6.9, 8 &  1.7, 8 \\
     7 &  [36.87, 16.16] &  6.8, 12 &  6.6, 32 &  6.5, 30  &  5.7, 25 &  6.9, 8 &  1.7, 7 \\
     8 &  [37.85, 13.31] &  6.8, 12 &  6.6, 32 &  6.5, 31  &  5.7, 25 &  6.9, 8 &  1.7, 8 \\
\tableline
\enddata
\tablecomments{$\Delta$t is the time duration spanned by each observing set in hours and
N$_\textrm{exp}$ is the total number of exposures for each night and field.}
\end{deluxetable}

\begin{deluxetable}{cccccc}
\tabletypesize{\scriptsize}
\tablecaption{Color observation on Oct 26 2016. \label{obs_log_1}}
\tablewidth{0pt}
\startdata \tableline\tableline
Field No. & $w_{P1}$ & $g_{P1}$ & $r_{P1}$ & $i_{P1}$ & $z_{P1}$ \\
         & N$_\textrm{exp}$ & $\Delta$t, N$_\textrm{exp}$ & $\Delta$t, N$_\textrm{exp}$ & $\Delta$t, N$_\textrm{exp}$ \\
\tableline
     1 &  6.8, 12 &  5.4, 3 &  5.4, 3 &  5.4, 3  &  5.4, 3  \\
     2 &  6.3, 11 &  5.4, 3 &  5.4, 3 &  5.4, 3  &  5.4, 3  \\
     3 &  6.8, 12 &  5.4, 3 &  5.4, 3 &  5.4, 3  &  5.4, 3  \\
     4 &  6.8, 12 &  5.4, 3 &  5.4, 3 &  5.4, 3  &  5.4, 3  \\
     5 &  6.8, 12 &  5.4, 3 &  5.4, 3 &  5.4, 3  &  5.4, 3  \\
     6 &  6.3, 11 &  5.4, 3 &  5.4, 3 &  5.4, 3  &  5.4, 3  \\
     7 &  6.8, 12 &  5.4, 3 &  5.4, 3 &  5.4, 3  &  5.4, 3  \\
     8 &  6.8, 12 &  5.4, 3 &  5.4, 3 &  5.4, 3  &  5.4, 3  \\
\tableline
\enddata
\tablecomments{$\Delta$t is the time duration spanned by each observing set in hours and
N$_\textrm{exp}$ is the total number of exposures for each night and field. The exposure time for
$g_{P1}$, $r_{P1}$, $i_{P1}$, $z_{P1}$, and $w_{P1}$ bands were 120, 120, 120, 180, and 60 seconds, respectively.
The observation sequence was $w_{P1} - g_{P1} - w_{P1} -r_{P1} - w_{P1} - i_{P1} - w_{P1} - z_{P1}$.}
\end{deluxetable}

\begin{deluxetable}{llrrrrrrrrrrrrrrrc}
\tabletypesize{\scriptsize} \setlength{\tabcolsep}{0.02in} \tablecaption{The 876 reliable rotation periods. \label{table_p}} \tablewidth{0pt} \tablehead{
\colhead{Obj ID} & \colhead{Designation} & \colhead{$a$} & \colhead{$e$} & \colhead{$i$} & \colhead{$\triangle$} & \colhead{$r$} &
\colhead{$\alpha$} & \colhead{$D$} & \colhead{$H$} & \colhead{$m$} & \colhead{Period} & \colhead{$\triangle$ m} & \colhead{$U$} & \colhead{$a^*$} & \colhead{$i-z$} & \colhead{Type} }
\startdata
\enddata
\tablecomments{Columns: compact id, designations, semi-major axis ($a$, AU), eccentricity ($e$,
degree), inclination ($i$, degree), diameter (D, km), mean heliocentric distance ($\triangle$, AU), mean geodesic
distance ($r$, AU), mean phase angle ($\alpha$, degree), diameter ($D$, km), absolute magnitude ($H$, mag), apparent magnitude ($m$, mag in $w_{P1}$), derived rotation period (hours), light-curve amplitude
(mag), rotation period quality code ($U$), SDSS color $a*$, SDSS color $i-z$, and Spectral type. The full table is available in the eletronic version.}
\tablenotetext{a}{Rotation period measurements available in the LCDB.} \tablenotetext{b}{Long-period objects with partial coverage on rotational phase.} \tablenotetext{c}{$WISE$/$NEOWISE$ diameter.}
\end{deluxetable}


\begin{deluxetable}{rllcccccrcrrrl}
\tabletypesize{\scriptsize} \tablecaption{Confirmed large SFRs to date.\label{sfr_tbl}} \tablewidth{0pt}
\tablehead{\colhead{} & \colhead{Asteroid} & \colhead{Tax.} & \colhead{Per.} & \colhead{$\Delta m$} & \colhead{Dia.} & \colhead{$H$} & \colhead{Coh.} & \colhead{$a$} & \colhead{$e$} & \colhead{$i$} & \colhead{$\Omega$} & \colhead{$\omega$} & \colhead{Ref.}\\	
\colhead{} & \colhead{} & \colhead{} & \colhead{(hours)} & \colhead{(mag)} & \colhead{(km)} & \colhead{(mag)} & \colhead{(Pa)} & \colhead{($au$)} & \colhead{} & \colhead{($^\circ$)} & \colhead{($^\circ$)} & \colhead{($^\circ$)} & \colhead{}	
	
	} \startdata


  (395043) & 2009 DY105 & V   & $1.23 \pm 0.00$ &  0.34    & $0.8 \pm 0.0$ & $17.1 \pm 0.1$ & 165.6         & 2.86 &  0.04 &  1.67 & 190.2 &  30.6 & This work \\
  (475443) & 2006 RG24  & C   & $0.99 \pm 0.00$ &  0.39    & $0.9 \pm 0.0$ & $18.9 \pm 0.1$ & 251.0         & 2.19 &  0.21 &  5.16 & 221.8 &  85.5 & This work \\
  (476215) & 2007 US107 & C   & $1.99 \pm 0.00$ &  0.36    & $1.5 \pm 0.1$ & $17.8 \pm 0.1$ & 117.8         & 2.71 &  0.09 &  1.50 & 332.0 &  46.0 & This work \\
           & 2016 UY68  & V   & $1.99 \pm 0.00$ &  0.49    & $0.4 \pm 0.0$ & $19.0 \pm 0.2$ &   9.1         & 2.62 &  0.23 &  3.51 &  35.9 &  61.1 & This work \\
           & 2016 UG94  & S   & $1.76 \pm 0.00$ &  0.47    & $0.5 \pm 0.0$ & $18.7 \pm 0.1$ &  28.6         & 2.57 &  0.21 &  3.63 &  40.1 & 322.4 & This work \\
           & 2016 UL98  & C   & $0.52 \pm 0.00$ &  0.45    & $0.7 \pm 0.0$ & $19.1 \pm 0.1$ & 622.2         & 2.59 &  0.25 &  3.34 &  26.7 & 314.8 & This work \\
           & 2016 UN129 & C   & $0.92 \pm 0.00$ &  0.55    & $1.2 \pm 0.1$ & $18.3 \pm 0.2$ & 573.3         & 2.57 &  0.06 &  4.87 &  22.0 & 160.0 & This work \\

  (455213) & 2001 OE84  & S   & $0.49 \pm 0.00$ &  0.5     & $0.7 \pm 0.1$ & $18.3 \pm 0.2$ & $\sim1500^b$  & 2.28 &  0.47 &  9.34 &  32.2 &   2.8 & \citet{Pravec2002}    \\
  (335433) & 2005 UW163 & V   & $1.29 \pm 0.01$ &  0.8     & $0.6 \pm 0.3$ & $17.7 \pm 0.3$ & $\sim200^b$   & 2.39 &  0.15 &  1.62 & 224.6 & 183.6 & \citet{Chang2014b}    \\
   (29075) & 1950 DA    & M   & $2.12 \pm 0.00$ &  0.2$^a$ & $1.3 \pm 0.1$ & $16.8 \pm 0.2$ & $64 \pm 20$   & 1.70 &  0.51 & 12.17 & 356.7 & 312.8 & \citet{Rozitis2014}   \\
   (60716) & 2000 GD65  & S   & $1.95 \pm 0.00$ &  0.3     & $2.0 \pm 0.6$ & $15.6 \pm 0.5$ & 150--450      & 2.42 &  0.10 &  3.17 &  42.1 & 162.4 & \citet{Polishook2016} \\
   (40511) & 1999 RE88  & S   & $1.96 \pm 0.01$ &  1.0     & $1.9 \pm 0.3$ & $16.4 \pm 0.3$ & $780 \pm 500$ & 2.38 &  0.17 &  2.04 & 341.6 & 279.8 & \citet{Chang2016}     \\
  (144977) & 2005 EC127 & V/A & $1.65 \pm 0.01$ &  0.5     & $0.6 \pm 0.1$ & $17.8 \pm 0.1$ & $47 \pm 30$   & 2.21 &  0.17 &  4.75 & 336.9 & 312.8 & \citet{Chang2017}     \\
\enddata
\tablecomments{The values of six known SFRs (i.e., after 2001 OE84) are adopted from \citet{Chang2017}}
\tablenotetext{a}{$\Delta m$ is adopted from \citet{Busch2007}.}
\tablenotetext{b}{The cohesion is adopted from \citet{Chang2016}.}
\end{deluxetable}

\clearpage
\begin{figure}
\includegraphics[angle=0,scale=.8]{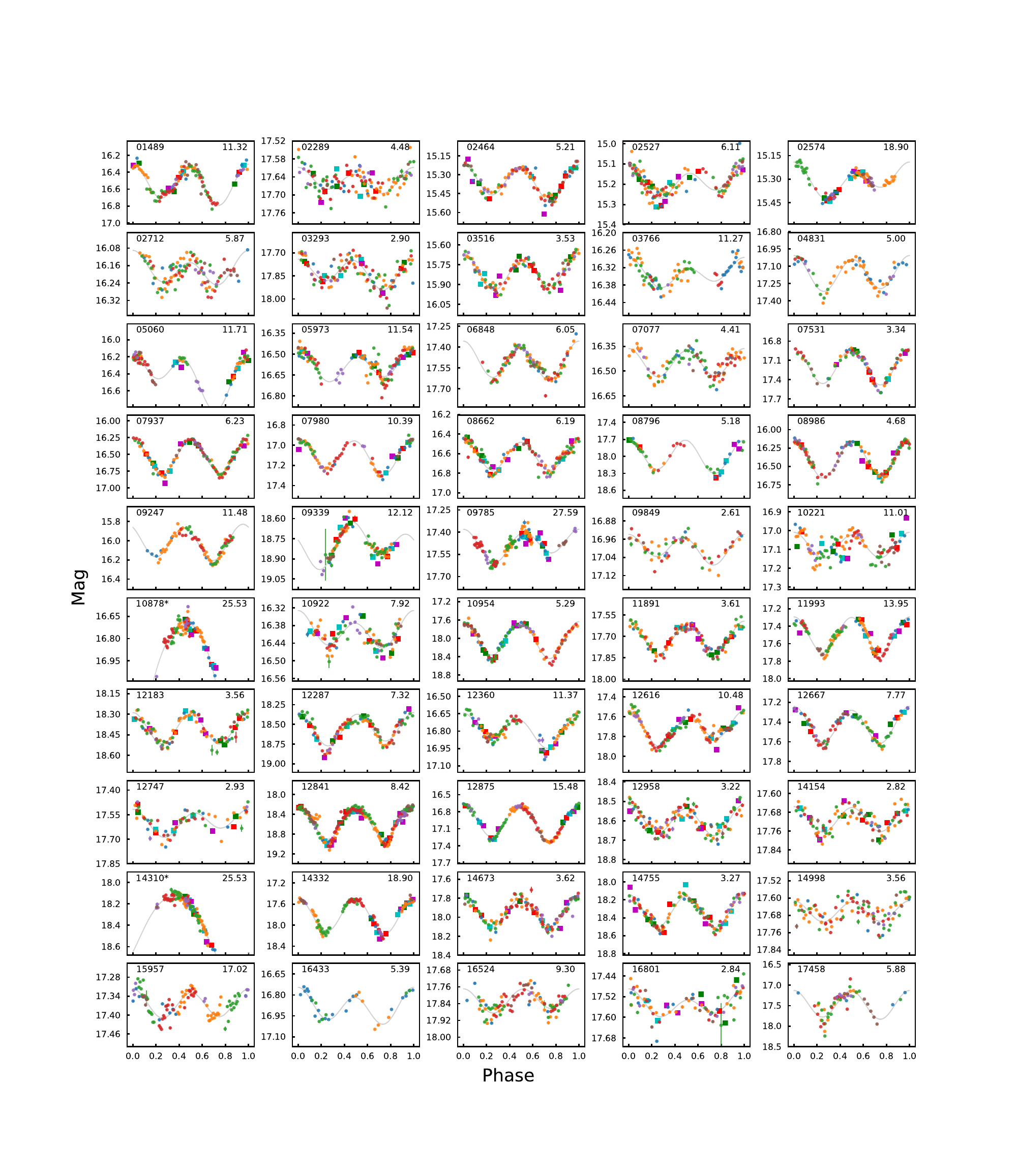}
\caption{Set of 50 folded lightcurves for the reliable rotation periods. Filled circles with different colors are data points in $w_{P1}$ band taken from different nights. The filled squares in green, red, canyon, and magenta are data points in $g_{P1}$, $r_{P1}$, $i_{P1}$, and $z_{P1}$ bands. The compact number/designation of asteroid is given on each plot along with its rotation period. Note that the data points in $g_{P1}$, $r_{P1}$, $i_{P1}$, and $z_{P1}$ bands are shifted to match the folded light curve in $w_{P1}$ band.}
\label{lightcurve00}
\end{figure}
\begin{figure}
\caption{Same as Fig.~\ref{lightcurve00} for other 50 reliable rotation periods.}
\label{lightcurve01}
\end{figure}
\begin{figure}
\caption{Same as Fig.~\ref{lightcurve00} for other 50 reliable rotation periods.}
\label{lightcurve02}
\end{figure}
\begin{figure}
\caption{Same as Fig.~\ref{lightcurve00} for other 50 reliable rotation periods.}
\label{lightcurve03}
\end{figure}
\begin{figure}
\caption{Same as Fig.~\ref{lightcurve00} for other 50 reliable rotation periods.}
\label{lightcurve04}
\end{figure}
\begin{figure}
\caption{Same as Fig.~\ref{lightcurve00} for other 50 reliable rotation periods.}
\label{lightcurve05}
\end{figure}

\clearpage
\begin{figure}
\caption{Same as Fig.~\ref{lightcurve00} for other 50 reliable rotation periods.}
\label{lightcurve06}
\end{figure}
\begin{figure}
\caption{Same as Fig.~\ref{lightcurve00} for other 50 reliable rotation periods.}
\label{lightcurve07}
\end{figure}
\begin{figure}
\caption{Same as Fig.~\ref{lightcurve00} for other 50 reliable rotation periods.}
\label{lightcurve08}
\end{figure}
\begin{figure}
\caption{Same as Fig.~\ref{lightcurve00} for other 50 reliable rotation periods.}
\label{lightcurve09}
\end{figure}
\begin{figure}
\caption{Same as Fig.~\ref{lightcurve00} for other 50 reliable rotation periods.}
\label{lightcurve10}
\end{figure}

\clearpage
\begin{figure}
\caption{Same as Fig.~\ref{lightcurve00} for other 50 reliable rotation periods.}
\label{lightcurve11}
\end{figure}
\begin{figure}
\caption{Same as Fig.~\ref{lightcurve00} for other 50 reliable rotation periods.}
\label{lightcurve12}
\end{figure}
\begin{figure}
\caption{Same as Fig.~\ref{lightcurve00} for other 50 reliable rotation periods.}
\label{lightcurve13}
\end{figure}
\begin{figure}
\caption{Same as Fig.~\ref{lightcurve00} for other 50 reliable rotation periods.}
\label{lightcurve14}
\end{figure}
\begin{figure}
\caption{Same as Fig.~\ref{lightcurve00} for other 50 reliable rotation periods.}
\label{lightcurve15}
\end{figure}
\begin{figure}
\caption{Same as Fig.~\ref{lightcurve00} for other 50 reliable rotation periods.}
\label{lightcurve16}
\end{figure}
\begin{figure}
\caption{Same as Fig.~\ref{lightcurve00} for other 26 reliable rotation periods.}
\label{lightcurve17}
\end{figure}

\end{document}